\documentclass[prc,twocolumn,epsfig,nofootinbib,floatfix,showpacs]{revtex4}
\usepackage{graphics}
\usepackage{epsfig}
\usepackage{amsfonts}
\usepackage{amsmath}
\usepackage{bm}% bold math

\usepackage{color}

\begin{document}
\title{
Microscopic analysis of low-energy spin and orbital
magnetic dipole excitations in deformed nuclei
}
\author{V.O. Nesterenko$^{1,2,3}$, P.I. Vishnevskiy $^{1,2,4}$,
J. Kvasil$^5$, A. Repko$^6$  and W. Kleinig$^1$}
\affiliation{$^1$
Laboratory of Theoretical Physics,
Joint Institute for Nuclear Research, Dubna, Moscow
region, 141980, Russia}
\email{nester@theor.jinr.ru}
\affiliation{$^2$
State University "Dubna", Dubna, Moscow Region, 141980, Russia}
\affiliation{$^3$
Moscow Institute of Physics and Technology,
Dolgoprudny, Moscow region, 141701, Russia}
\affiliation{$^4$
 Institute of Nuclear Physic, Almaty, 050032, Kazakhstan}
\affiliation{$^5$
Institute of Particle and Nuclear Physics, Charles
University, CZ-18000, Praha 8, Czech Republic}
\affiliation{$^6$
Institute of Physics, Slovak Academy of Sciences, 84511, Bratislava, Slovakia}

\date{\today}

\begin{abstract}
 A low-energy magnetic dipole $(M1)$ spin-scissors resonance (SSR) located just
 below the ordinary orbital scissors resonance  (OSR) was recently predicted
 in deformed nuclei within the Wigner Function Moments (WFM) approach. We analyze
 this prediction using fully self-consistent Skyrme Quasiparticle Random Phase
 Approximation (QRPA) method. Skyrme forces SkM*, SVbas and SG2 are implemented
 to explore SSR and OSR in $^{160,162,164}$Dy and $^{232}$Th. Accuracy of the method
 is justified by a good description of M1 spin-flip giant resonance. The calculations show
 that isotopes $^{160,162,164}$Dy indeed have at 1.5-2.4 MeV (below OSR)
 $I^{\pi}K=1^+1$ states  with a large $M1$ spin strength ($K$ is the projection
 of the total nuclear moment to the symmetry z-axis). These states are almost fully
exhausted by $pp[411\uparrow, 411\downarrow]$ and $nn[521\uparrow, 521\downarrow]$
spin-flip configurations corresponding  to $pp[2d_{3/2}, 2d_{5/2}]$ and
$nn[2f_{5/2}, 2f_{7/2}]$ structures in the spherical limit. So the predicted SSR
is actually reduced to low-orbital (l=2,3) spin-flip states. Following
our analysis and in contradiction with WFM spin-scissors picture, deformation
is not the principle origin of the low-energy spin $M1$ states but only a factor
affecting their features. The spin  and orbital strengths are generally mixed and
exhibit the interference: weak destructive in SSR range and strong constructive in OSR
range. In $^{232}$Th, the $M1$ spin strength is found very small. Two groups
of  $I^{\pi}=1^+$ states observed experimentally at 2.4-4 MeV in $^{160,162,164}$Dy  and
at 2-4 MeV in $^{232}$Th are mainly explained by fragmentation of the orbital strength.
%than by the occurrence of spin-flip states.
Distributions of nuclear currents in QRPA states partly correspond to the isovector
orbital-scissors flow but not to spin-scissors one.

\end{abstract}

\maketitle
\section{Introduction}
\label{intro}

Magnetic dipole excitations in nuclei provide important information on
the nuclear spin and orbital magnetism  \cite{Har01,Hei10}.
For a long time, these excitations  were mainly represented by $M1(K=1)$ spin-flip
giant resonance located at the energy E$\approx 41 A^{-1/3}$ MeV
\cite{Har01,Hei10} and low-energy $M1$ OSR
with excitation energy E$\approx 66 \delta A^{-1/3}$ MeV \cite{Hei10}
where $\delta$ is the parameter of nuclear axial quadrupole deformation.
Both resonances are isovector and characterized by
enhanced $M1(\Delta K)$ transitions to the ground state.

The spin-flip resonance is produced by particle-hole spin-flip transitions
between spin-orbit partners in the proton and neutron single-particle spectra.
This resonance is related to  {\it spin} nuclear magnetic properties and
it exists in both spherical and deformed nuclei \cite{Har01,Hei10}. 
The spin-flip resonance was widely applied to test a spin channel in various
self-consistent approaches (Skyrme, Gogny and relativistic)
\cite{Hei10,Les07,Ves_PRC09,Nes_JPG10,Gor_PRC16,Paar_20,Tse_PRC19} and to check
tensor forces \cite{Les07,Ves_PRC09,Li_PRC09} and spin-orbit
interaction \cite{Les07,Ves_PRC09,Nes_JPG10,Paar_20}.

 \begin{figure} %1
\centering
\includegraphics[width=6cm]{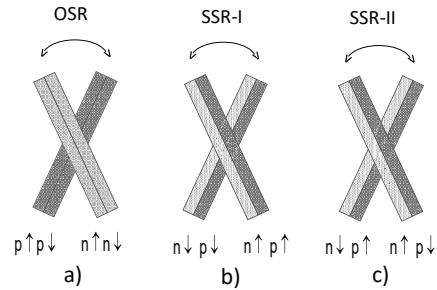}
\caption{The schemes for the members of the scissors triple \cite{Bal_arxiv19}:
OSR (a), SSR-I (b) and  SSR-II (c). The neutron (proton) axially deformed fractions
are shown by light (dark) bars. The spin direction of nucleons is indicated by arrows.
Each mode in the triple exhibits scissors-like oscillations of two blades:
neutrons vs protons in OSR, spin-up vs spin-down nucleons in SSR-I
(spins of neutrons and protons in each blade have the same direction), and
SSR-II where neutron and proton spins in each blade have opposite directions.}
\label{fig1}
\end{figure}

OSR is macroscopically treated as scissors-like out-of-phase
oscillations of proton and neutron deformed subsystems, see Fig.~1a. This
isovector resonance can exist only in deformed nuclei. It
represents a remarkable example of a nuclear {\it orbital} magnetism.
OSR was predicted in the two-rotor model \cite{Iud78,Hil84} and then experimentally
observed in $(e,e')$ reaction \cite{Boh84}. OSR demonstrates some  specific
features: linear and square deformation laws for its energy
and strength, respectively \cite{Zie_PRL90,Iud_PLB93}. Various properties of OSR
are outlined in reviews \cite{Hei10,Iud_PN97,Iud_NC00}.
OSR is a kind of mixed-symmetry state \cite{Die_PPNP83,Iud_Sto_PRC02,Piet_MSS}.
Recent studies of OSR can be found elsewhere, see e.g.
\cite{End_analysis_2005,Pai16,Gut12}.

A decade ago, E.B. Balbutsev, I.V. Molodtsova, and P. Schuck
have predicted (within the WFM method) that OSR should be
supplemented by a low-energy {\it spin} scissor mode (SSR) \cite{Bal_NPA11}.
Further WFM calculations with inclusion of the pairing
\cite{Bal_PRC15,Bal_PRC18,Mol_EPJC18} and isoscalar-isovector
coupling in the  residual interaction
\cite{Bal_EPJC18,Bal_arxiv19,Bal_PAN20} have shown that SSR should have two branches,
(see Fig.~1b,c) lying  below OSR. Thus altogether the nuclear scissors mode
should be a triplet: OSR + two SSR branches. All the scissors states
should demonstrate significant $M1(\Delta K=1)$ transitions to the ground state.

Following the WFM calculations, SSR should exist in medium and heavy
axial deformed nuclei, typically at the excitation energy $E <$  2.7 MeV, i.e.
just below OSR \cite{Bal_PRC18,Mol_EPJC18,Bal_EPJC18,Bal_arxiv19,Bal_PAN20}.
Many $I^{\pi}=1^+ $ states at $E <$  2.7 MeV were already observed in
rare-earth and actinide nuclei, see e.g.
\cite{Wessel_PLB88,Mar_exp_NRF_2005,Val_exp_2015,Ren_exp_Oslo_2018,Ade_232Th}. However,
they  are usually not included in the experimental OSR systematics and their
origin is still rather unclear. The  prediction of SSR suggests an explanation
for these states. Following the detailed WFM analysis for $^{160,162,164}$Dy,
$^{232}$Th and $^{236,238}$U \cite{Mol_EPJC18,Bal_EPJC18,Bal_arxiv19,Bal_PAN20}, the
nuclei $^{164}$Dy and  $^{232}$Th are the most promising candidates for SSR.
Low-energy $1^+$ states in these nuclei form two
distinctive groups which might be attributed to SSR and OSR.

\begin{figure} %2
\centering
\includegraphics[width=8cm]{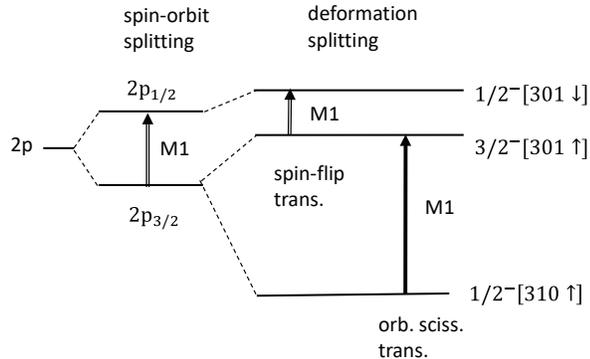}
\caption{A scheme of single-particle levels for $2p$ subshell in spherical
(left) and  deformed (right) cases. The scheme corresponds to the proton $2p$ subshell
in $^{162}$Dy, calculated with the Skyrme force SG2. Spin-flip and orbital
scissors $M1$ transitions are exhibited by empty and filled arrows, respectively.
In the deformed case, the levels are denoted by Nilsson asymptotic quantum numbers
\cite{Sol,Nil}, the arrows indicate spin direction.}
\label{fig2}
\end{figure}

The aim of the present paper is to scrutinize the WFM prediction of SSR
from the microscopic viewpoint. It is well known that both
orbital and spin-flip M1 transitions can be explained using
single-particle schemes \cite{Har01,RSbook}. An example of such scheme
for $2p$-subshell is shown in Fig.~2. This is a fraction of the proton scheme
in $^{162}$Dy, calculated with Skyrme parametrization SG2 \cite{SG2}.
The computed equilibrium axial quadrupole deformation is $\beta_2$=0.346.
The left part of the figure shows the splitting of  $2p$-subshell
into $2p_{1/2}$ and $2p_{3/2}$ levels due to spin-orbit interaction.
Already in this spherical case, a spin-flip $M1$ transition between the levels
is possible. The large deformation significantly splits the level $2p_{3/2}$
and upshifts the level $2p_{1/2}$ (right part of Fig. 2). In this case,
two $M1(\Delta K=1)$ transitions are possible: spin-flip
$3/2^-[301\uparrow] \to 1/2^-[301\downarrow]$ and orbital
$1/2^-[310\uparrow] \to  3/2^-[301\uparrow]$. The former connects
the spin-orbit partners, the latter relates the levels arising due
to deformation splitting. So we get two natural candidates
for SSR and OSR. Because of the large deformation splitting, the
orbital transition has a larger energy than the spin-flip
one. So SSR should lie lower by energy than OSR.

As seen in Fig. 1(b,c), neutrons  and protons in the left and
right scissors blades have opposite spin directions. Perhaps,
the predicted SSR can be somehow related to spin-flip excitations
in neutron and proton spectra. This point is yet unclear (see
discussion in Appendix B). What is important, Fig. 2 clearly shows
that nuclear deformation is not the primary origin of low-energy
spin-flip states (though it can significantly affect their features).
This means that WFM interpretation of low-energy spin states in
terms of {\it deformation-induced scissors} oscillations is questionable.

The main aim of the present study is to show that the predicted low-energy
spin states are ordinary spin-flip excitations and the available experimental data
can be explained by the fragmentation of spin-flip and orbital M1 strength.
Our analysis is performed  for axially deformed nuclei $^{160,162,164}$Dy
and $^{232}$Th. As mentioned above, two of these nuclei, $^{164}$Dy and $^{232}$Th,
are considered by WFM as promising candidates for SSR. The calculations are
performed using fully self-consistent QRPA
\cite{Ben_RMP03,Stone_PPNP07,Rep_arxiv,Rep_EPJA17,Rep_sePRC19,Kva_seEPJA19}
with the Skyrme forces SG2 \cite{SG2}, SkM* \cite{SkM*}, and SVbas \cite{SVbas}.
As shown below, the spin and orbital low-energy $M1$ excitations are strongly
mixed.  So we will analyze both SSR and OSR. To demonstrate accuracy of
our calculations, we will also present results for $M1$ spin-flip giant resonance.

The paper is organized as follows. In Sec. II, the calculation scheme is outlined.
In Sec. III, results of the calculations are discussed. In particular, flows
the nuclear currents are exhibited. In Sec. IV, the conclusions are done.
In Appendix A, a description of the $M1$ spin-flip giant resonance is illustrated.
In Appendix B, some important aspects of WFM/QRPA  comparison are commented.
In Appendix C, expressions for the orbit and spin transition matrix elements are given.

\section{Calculation scheme}

The calculations are performed within the Skyrme QRPA model
\cite{Ben_RMP03,Stone_PPNP07,Rep_arxiv,Rep_EPJA17,Rep_sePRC19,Kva_seEPJA19}.
%\cite{Rep_arxiv,Rep_EPJA17,Rep_sePRC19,Kva_seEPJA19} based on the Skyrme 
%functionaln \cite{Ben_RMP03,Stone_PPNP07}.
The model is fully self-consistent, i.e.: i) both mean field and residual
interaction are derived from the initial Skyrme functional, ii) the residual
interaction takes into account all the terms of the Skyrme functional and Coulomb
(direct and exchange) parts, iii) both particle-hole  and particle-particle
channels are included \cite{Rep_EPJA17}.  Spurious admixtures caused by violation
of the rotational invariance are removed using the technique SEBRPA (spuriosity 
extracted before RPA) \cite{Kva_seEPJA19}.

A representative set of Skyrme forces is used. We employ
the standard force SkM* \cite{SkM*}, the recently developed force SVbas
\cite{SVbas}, and the force SG2 \cite{SG2} which is often used
in analysis of magnetic excitations,
see e.g. \cite{Ves_PRC09,Nes_JPG10,Sarr96}. As seen from Table \ref{tab-1},
these forces have different isoscalar $b_4$ and isovector $b_4'$ parameters
of the spin-orbit terms in the Skyrme functionals (see definitions of the 
parameters in Refs. \cite{Ves_PRC09,Stone_PPNP07}). In SkM* and SG2,
the usual convention $b_4=b'_4$ is used while in SVbas a separate tuning
of $b_4$ and $b_4'$ is done. All three Skyrme forces reproduce, though with
different degrees of accuracy, a two-hump structure of M1 spin-flip
giant resonance in deformed nuclei \cite{Ves_PRC09,Nes_JPG10}.
As shown in Appendix A, SVbas and especially SG2 give a nice description
of this resonance. So these two Skyrme forces can be considered as
the most relevant for the present study.

The nuclear mean field and pairing are computed with the code SKYAX
\cite{SKYAX} using a two-dimensional grid in cylindrical coordinates.
The calculation box extends up to three times the nuclear radii, the
grid step is 0.4 fm. The axial quadrupole equilibrium deformation is
obtained by minimization of
the energy of the system. As seen from Table \ref{tab-2}, the obtained values
of the deformation  parameter $\beta$  are in a good agreement with
the experimental data \cite{database}, especially for SVbas. All the forces
reproduce a grow of the deformation from $^{160}$Dy to $^{164}$Dy.

\begin{table} % Table I
\centering
\caption{Isoscalar effective mass $m^*_0$,  isoscalar and isovector spin-orbit
parameters $b_4$ and $b'_4$, proton and neutron pairing constants  $G_p$ and
$G_n$, and the type of pairing in Skyrme forces SkM*, SVbas, and SG2.}
\label{tab-1}       % Give a unique label
\begin{tabular}{|c|c|c|c|c|c|c|}
\hline
 force       & $m^*_0$  & $b_4$ & $b'_4$ & $G_p$ & $G_n$ & pairing \\
 & &  MeV $\rm{fm}^5$      &  MeV $\rm{fm}^5$  &  MeV $\rm{fm}^3$ &  MeV $\rm{fm}^3$ &  \\
\hline
SkM* & \; 0.79 & 65.0 & 65.0 & 279.08 & 258.96 & volume\\
\hline
SVbas  & \; 0.90 & 62.32 & 34.11 & 674.62 & 606.90 & surface   \\
\hline
 SG2  & \; 0.79 & 52.5 & 52.5 & 296.76 & 259.58 & volume  \\
\hline
\end{tabular}
\end{table}
\begin{table} % Table II
\centering
\caption{Calculated parameters $\beta$ of the equilibrium axial quadrupole
deformation vs the experimental values  \cite{database}.}
\label{tab-2}       % Give a unique label
\begin{tabular}{|c|c|c|c|c|}
\hline
Nucleus & \multicolumn{4}{c|}{$\beta$}\\
\hline
        &  SkM* & SVbas & SG2 & Exper. \\
\hline
$^{160}$Dy  & \; 0.339& 0.331 & 0.339 & 0.334 (2)  \\
\hline
$^{162}$Dy  & \; 0.351 & 0.345 & 0.346  &  0.341(3)  \\
\hline
$^{164}$Dy  & \; 0.354 & 0.348 & 0.352  & 0.349(3)\\
\hline
$^{232}$Th &  \; 0.256 & \; 0.247 & \: 0.238 & 0.248 (6) \\
\hline
\end{tabular}
\end{table}

Pairing is described by the zero-range pairing interaction \cite{Be00}
\begin{equation}
  V^{q}_{\rm pair}(\bold{r},\bold{r}') =  G_{q} \Big[ 1 - \eta \:
\Big(\frac{\rho(\bold{r})}{\rho_{\rm pair}}\Big)\Big]\delta(\bold{r}-\bold{r}')
\label{Vpair}
\end{equation}
where $G_{q}$ are proton ($q=p$) and neutron ($q=n$) pairing strength constants.
They are fitted to reproduce empirical pairing gaps obtained by the five-point
formula along selected isotopic and isotonic chains \cite{G_Rein}. The values
of $G_{q}$ are shown in Table \ref{tab-1}.
Further, $\rho(\bold{r})=\rho_p(\bold{r})+\rho_n(\bold{r})$ is the sum of
proton and neutron densities. We get so-called volume pairing for
$\eta$=0  and density-dependent surface pairing for $\eta$=1.
As indicated in Table \ref{tab-1}, the former is used in SkM*
and SG2, and the latter is exploited in SVbas. In the latter case,
we use SVbas parameter $\rho_{\rm pair}$=0.2011 ${\rm fm}^{-3}$.
Pairing correlations are included at the level of the iterative HF-BCS
(Hartree-Fock plus Bardeen-Cooper-Schrieffer) method \cite{Rep_EPJA17}.
To cope with the divergent character of zero-range pairing forces,
energy-dependent cut-off factors are used \cite{Rep_EPJA17,Be00}.

\begin{table} % Table III
\centering
\caption{Proton and neutron pairing gaps $\Delta_p$ and $\Delta_n$ and
energy of $2^+_1$ state of the ground-state rotational band, calculated
in $^{162}$Dy and $^{232}$Th with Skyrme forces SkM*, SVbas, and SG2.
The experimental data for the energy $E_{2^+_1}$ are taken from database
\cite{database}.}
\label{tab-3}       % Give a unique label
\begin{tabular}{|c|c|c|c|c|c|}
\hline
 Nucleus &        & SkM*  & SVbas & SG2 & exper. \\
\hline
           & $\Delta_p$ [MeV] & 0.55 & 0.69 & 0.72 & \\
$^{162}$Dy & $\Delta_n$ [MeV] & 0.62 & 0.95 & 0.87 & \\
            & $E_{2^+_1}$ [keV] & 67.9 & 92.7 & 88.8 & 80.7 \\
\hline
           & $\Delta_p$ [MeV] & 0.53 & 0.61 & 0.75 & \\
$^{232}$Th & $\Delta_n$ [MeV] & 0.54 & 0.80 & 0.78 & \\
            & $E_{2^+_1}$ [keV] & 41.2 & 57.1 & 63.0 & 49.4 \\
\hline
\end{tabular}
\end{table}

Table  \ref{tab-3} shows the calculated averaged proton and neutron
pairing gaps $\Delta_p$ and $\Delta_n$ (defined in Eq. (30) of Ref.
\cite{Be00}) in $^{162}$Dy and $^{232}$Th.
Also we exhibit the energies $E_{2^+_1}=3\hbar^2/\mathcal{J}$ (with
$\mathcal{J}$ being the
nuclear moment of inertia) of  $I^{\pi}=2^+$ state in ground-state
rotational band.
These energies  are sensitive to both deformation and pairing.
As seen from Table  \ref{tab-3}, SkM* underestimates while SVbas and SG2
somewhat overestimate the experimental  $E_{2^+_1}$-values.

In our calculations, QRPA is implemented in the matrix form. A large
configuration space is used. The single-particle spectrum extends
from the bottom of the potential well up to 30 MeV. For example,
in SG2 calculations for  $^{162}$Dy,  691 proton and 800 neutron
single-particle levels are used.  The two-quasiparticle (2qp) basis
in QRPA calculation for $K^{\pi}=1^+$ states includes 5270 proton
and 9527 neutron configurations. We do not consider $K^{\pi}=0^+$
excitations since it is well known \cite{Har01,Hei10,Iud_PN97,Iud_NC00}
that $M1$ spin-flip and orbital-scissors modes are characterized
by strong $M1(\Delta K=1)$ transitions to the ground state.

Reduced probability for $M1$ transitions
from the ground state $|0\rangle$ with $I^{\pi}K=0^+0$ to
the excited QRPA state $|\nu\rangle$ with $I^{\pi}K=1^+1$ reads
\begin{equation}
\label{eq:BM11}
B_{\nu}(M1)=2|\:\langle\nu|\:\hat{\Gamma}(M11)\:|0\rangle \:|^2 .
\end{equation}
The coefficient 2 means that contributions of both projections $K$=1
and -1 are taken into account.  The transition operator has the form
\begin{equation}
\label{eq:M1}
 {\hat \Gamma}(M11) =
 \mu_N \sqrt{\frac{3}{4\pi}}\sum_{q = p,n}
 %\sum_{k \epsilon q}
[g^{q}_s {\hat s}(\mu=1) + g^{q}_l {\hat l} (\mu=1)]
\end{equation}
where $\mu_N$ is the nuclear magneton, ${\hat{s}}(\mu=1)$ and ${\hat{l}}(\mu=1)$ are
$\mu$=1 projections of the standard spin and orbital operators,
$g^{q}_s$ and $g^{q}_l$
are spin and orbital gyromagnetic factors. We use the quenched spin g-factors
$g^{q}_s = \eta \bar{g}^{q}_s$ where  $\bar{g}^{p}_s$ = 5.58  and  $\bar{g}^{n}_s$ =-3.82
are bare proton and neutron g-factors and $\eta$=0.7 is the
quenching parameter \cite{Har01}. The orbital g-factors are  $g^{p}_l$ = 1
and  $g^{n}_l$ = 0. In what follows, we consider three
cases: spin ($g^{q}_l=0$), orbital ($g^{q}_s=0$), and total (when both
spin and orbital transitions are taken into account). The expressions for
orbital and spin $M1$ matrix elements are given in the Appendix C.

In deformed nuclei, electric and magnetic states with the same $K^{\pi}$
are mixed \cite{Har01,RSbook,Iud_PN97,Piet_PRC95}. In our case
of $K^{\pi}=1^+$ states, the magnetic dipole  $M1(K=1)$ and electric quadrupole
$E2(K=1)$ modes can be mixed. To estimate this mixing, we calculate
reduced probability of E2 transitions $0^+0 \to 2^+1$:
\begin{equation}
\label{BE21}
B_{\nu}(E2)=2|\:\langle\nu|\:\hat{\Gamma}(E21)\:|0\rangle \:|^2
\end{equation}
with the proton transition operator
\begin{equation}
\label{eq:E21}
 {\hat \Gamma}(E21) =
  e r^2 Y_{21}(\theta,\phi)
  %\sum_{q \epsilon p}
  %e_{\text{eff}}^q
%\sum_{k \epsilon q}
\end{equation}
where $Y_{21}(\theta,\phi)$ is the spherical harmonic.

We also calculate the current transition densities (CTD)
\begin{equation}
\delta \bold {j}_{\nu}(\bold{r}) = \langle \nu| \hat{\bold j}|0\rangle (\bold{r})
\label{CTD}
\end{equation}
for the convective nuclear current
\begin{equation}
\label{j_con}
\hat{\bold j} (\bold r)= -i \frac{e\hbar}{2m} \sum_{q =n,p}e_{\text{eff}}^q
\sum_{k \epsilon q}(\delta({\bold r} - {\bold r}_k) {\bold \nabla}_k
+ {\bold \nabla}_k \delta({\bold r} - {\bold r}_k)) .
\end{equation}
Here $e_{\text{eff}}^q$ are the effective charges. They are
$e_{\text{eff}}^p$=1 and $e_{\text{eff}}^n$=0 for the proton current,
$e_{\text{eff}}^p$=0 and $e_{\text{eff}}^n$=1 for the neutron current,
$e_{\text{eff}}^p=e_{\text{eff}}^n$=1 for isoscalar current and
$e_{\text{eff}}^p=-e_{\text{eff}}^n$=1 for isovector current.

Beside, we calculate the separate spin-up and spin-down parts of CTD (\ref{CTD}).
For this aim, the wave function of the QRPA state
$|\nu \rangle $ is projected
to the proper spin direction using spinor structure of the involved
single-particle wave functions in cylindric coordinates, see Eqs.
(\ref{eq:psi_i_spin})-(\ref{eq:psi_i_bar_spin}) in Appendix C.

\begin{figure*} %3
\centering
\includegraphics[width=11cm]{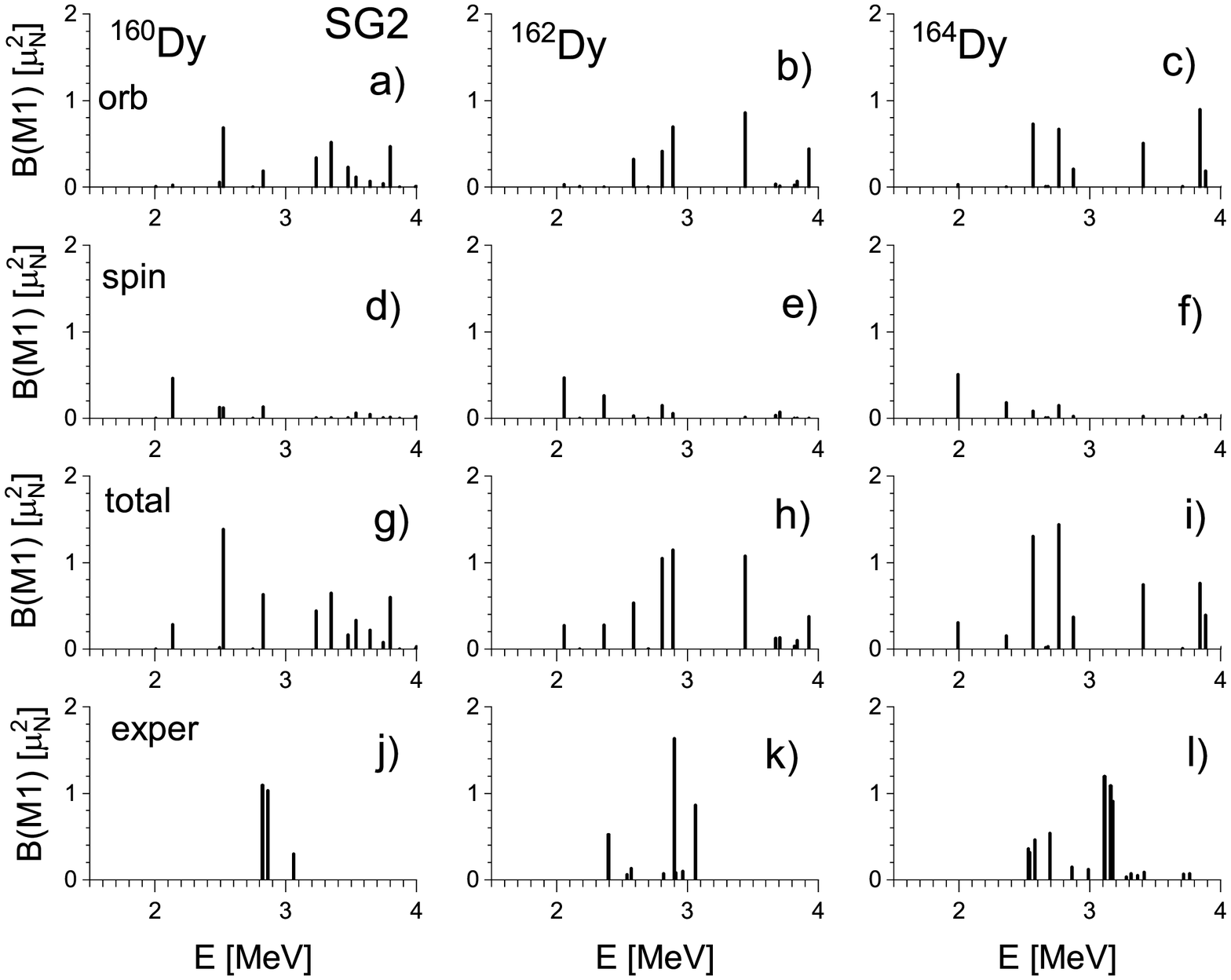}
\caption{Orbital (a,b,c), spin (d,e,f) and total (g,h,i)  low-energy M1
strength in $^{160,162,164}$Dy, calculated in QRPA with Skyrme force SG2.
In the bottom panels, the experimental M1 strength for $^{160}$Dy
\cite{Wessel_PLB88} and $^{162,164}$Dy \cite{Mar_exp_NRF_2005} is shown.}
\label{fig3_Dy}
\end{figure*}

\section{Results and discussion}

\subsection{$M1$ strength  in  $^{160,162,164}$Dy}

In Figure 3, we compare calculated orbital, spin and total M1
strengths (\ref{eq:BM11}) in $^{160,162,164}$Dy with experimental
data from the nuclear resonance fluorescence
(NRF) reaction, see Refs.  \cite{Wessel_PLB88} for $^{160}$Dy and
\cite{Mar_exp_NRF_2005} for $^{162,164}$Dy. QRPA results are obtained
for the force SG2. Following the discussion in Sec. II and results
for the spin-flip $M1$ giant resonance in Appendix A,
this force seems to be the most relevant  for our analysis.

The plots (a-c) of the figure show that $M1$ strength above 2.4 MeV is mainly orbital.
This strength constitutes the OSR. Instead, a few states at E $<$ 2.4 MeV
exhibit a noticeable  spin strength, see plots (d-f). Following prediction
\cite{Bal_PRC18,Mol_EPJC18,Bal_EPJC18,Bal_arxiv19,Bal_PAN20},
these states are candidates for SSR. Comparing spin and orbital strengths
with the total one (plots (g-i)), we see that spin and orbital modes have
a strong interference, both destructive and constructive. These results
take place for all three Dy isotopes.
\begin{figure*} %4
\centering
\includegraphics[width=11cm]{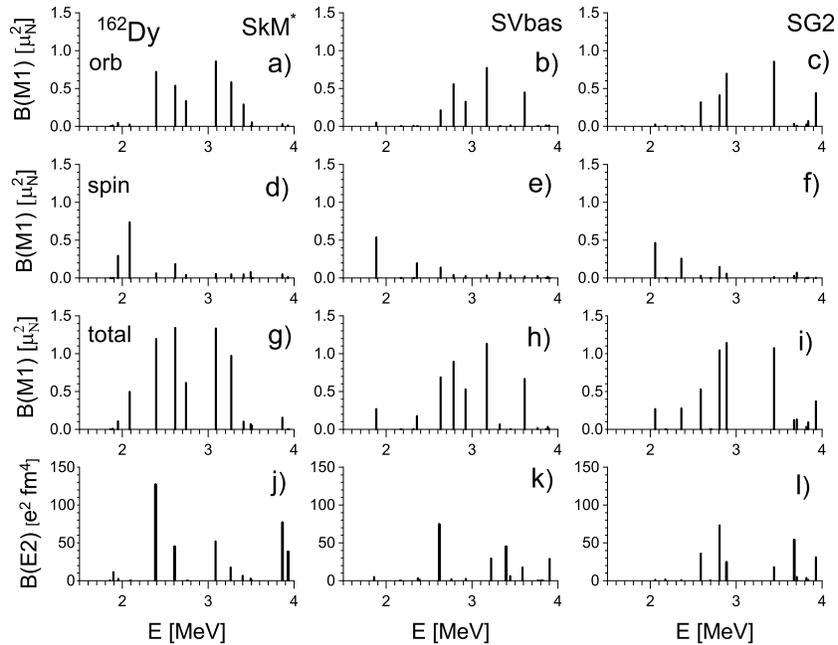}
\caption{Orbital (a,b,c), spin (d,e,f) and total (g,h,i)
low-energy M1 strength in $^{162}$Dy, calculated in QRPA
with Skyrme forces SkM* (left), SVbas (middle) and SG2 (right).
In the bottom panels, the quadrupole E2 strength is shown.}
\label{fig4_162Dy}
\end{figure*}

\begin{table*} % Table IV
%\centering
\caption{
The calculated orbital, spin and total strengths $\sum B(M1)$ (in $\mu^2_N)$
in $^{160,162,164}$Dy, summed at SSR (0-2.4 MeV), OSR (2.4-4 MeV) and total (0-4 MeV)
energy ranges as compared with experimental
data  for $^{160}$Dy \cite{Wessel_PLB88} and $^{162,164}$Dy \cite{Mar_exp_NRF_2005}.
For each energy range, the interference factors $R$ are shown.}
\label{tab-4}       % Give a unique label
\begin{tabular}{|c|c|ccc|c|ccc|c|cccc|c|}
\hline
Nucleus & Force & \multicolumn{4}{c|}{0-2.4 MeV}
                & \multicolumn{4}{c|}{2.4-4 MeV}
                & \multicolumn{5}{c|}{0-4 MeV}
%  && \multicolumn{8}{c|}{$\sum B(M11) \; [\mu^2_N]$}
\\
\hline
&  & \multicolumn{3}{c|}{$\sum B(M1)$} & R
                & \multicolumn{3}{c|}{$\sum B(M1)$} & R
                & \multicolumn{4}{c|}{$\sum B(M1)$} & R
\\
\hline
 &   & orb & spin & total & & orb & spin & total &  & orb & spin & total & exp  &
\\
\hline
            & \; SkM* &
\; 0.52 \; & \; 0.96 \; & \; 1.32 \; & 0.89 &
\; 2.79 \; & \; 0.55 \; & \; 4.85 \; & 1.45 &
\; 3.31 \; & \; 1.51 \; & \; 6.16 \; & \; & 1.28
\\
$^{160}$Dy &\; SVbas &
\; 0.05 \; & \; 0.49 \; & \; 0.23 \; &  0.43 &
\; 2.15 \; & \; 0.51 \; & \; 3.80 \; &  1.43 &
\; 2.20 \; & \; 1.00 \; & \; 4.03 \; & \; 2.42 \; & 1.26
\\
           & \; SG2 &
0.03 & 0.46 & 0.28 &  0.57 &
\; 2.69 \; & \; 0.54 \; & 4.53 &  1.40 &
\; 2.72 \; & \; 1.00 \; & 4.81 &  \; & 1.29
\\
\hline
           & \; SkM* &
 0.80 & 1.09 & 1.80 &  0.95 &
 2.69 & 0.51 & 4.63 &  1.45 &
 3.49 &  1.60 & 6.44 & \; & 1.27
 \\
$^{162}$Dy & \; SVbas &
0.06 & 0.73 & 0.45 &  0.57 &
2.35 & 0.40 & 4.04 &  1.47 &
2.41 & 1.14 & 4.49 & 3.45 & 1.26
\\
           & \; SG2   &
0.03 & 0.72 & 0.55 & 0.73 &
2.85 & 0.35 & 4.54 & 1.42 &
2.88 & 1.07 & 5.09 & \; & 1.29
\\
\hline
           & \; SkM* &
0.96 & 1.09 & 2.11 &   1.03 &
2.18 & 0.40 & 3.94 &   1.53 &
3.14 & 1.49 & 6.05 & \;  & 1.31
 \\
$^{164}$Dy & \; SVbas &
0.06 &  0.63 & 0.32 &  0.47 &
2.52 & 0.50 & 4.37 &  1.45 &
2.57 & 1.13 & 4.69 & 6.17 & 1.27
\\
            & \; SG2   &
0.03 & 0.68 & 0.45  &  0.63 &
3.20 & 0.35 & 5.05 &  1.42 &
3.23 & 1.03 & 5.50 & \; & 1.29
\\
\hline
\end{tabular}
\\
\end{table*}

Figure~3 shows that NRF data \cite{Wessel_PLB88,Mar_exp_NRF_2005} do not
give $I^{\pi}=1^+$ states at $E < $ 2.39 MeV. As discussed in Refs.
\cite{Ren_exp_Oslo_2018}, this may be caused by troubles of traditional
NRF experiments  to separate transitions in this energy range from a
sizable background.  The early data for $^{160 }$Dy  \cite{Wessel_PLB88}
give $1^+$ states only for $E >$ 2.8  MeV,
though the level list in database \cite{database} suggests many
candidates for $1^+$ states at lower excitation energies.

In $^{162,164}$Dy, NRF data \cite{Mar_exp_NRF_2005} give two groups
of $1^+$ states located above and below 2.7 MeV.
The former group is usually treated as OSR. The latter is treated by WFM as
SSR  \cite{Bal_PRC18,Mol_EPJC18,Bal_EPJC18,Bal_arxiv19,Bal_PAN20}.
Note that low-energy
groups of $1^+$ states were earlier observed in various rare-earth nuclei
\cite{Wessel_PLB88}. Recent Oslo $(\gamma,n)$  experiments  \cite{Ren_exp_Oslo_2018}
show that, in $^{164}$Dy, $40-60 \%$ of $M1$ strength at energy range 0-4 MeV is
located below 2.7 MeV. Moreover, in this nucleus the total measured  $M1$ strength
at 0-4 MeV achieves 6.17 $\mu^2_N$  \cite{Mar_exp_NRF_2005} which substantially exceeds
the values 3-4 $\mu^2_N$ typical for OSR in well-deformed rare-earth nuclei.
This observation was treated by WFM as a clear signature of SSR in $^{164}$Dy
 \cite{Bal_PRC18,Mol_EPJC18,Bal_EPJC18,Bal_arxiv19,Bal_PAN20}.
However, following our results in Fig. 3, the states at 2.4-2.7 MeV give mainly
orbital $M11$ transitions and so should also belong to OSR. They are omitted in
OSR systematics with the lower boundary 2.7 MeV \cite{Piet_PRC95} but taken into
account for the lower boundary 2.5 MeV \cite{End_analysis_2005}.
So, by our opinion, the data of Oslo group cannot be considered as
the argument in favor of SSR.

In Figure~4, we demonstrate the distribution of M1 strength in $^{162}$Dy,
calculated with the forces SkM*, SVbas, and SG2. It is seen
that, despite some deviations in details, all these three forces
give qualitatively similar results. In all cases, there is the
range  0-2.4 MeV with an essential spin strength and the range 2.4-4.0
MeV with a dominant orbital strength. Fig. 4 also demonstrates $E2$ strength
(\ref{BE21}) for
the same $K^{\pi}=1^+$ states. This strength is large at 2.6-4.0 MeV
and negligible at 0-2.6 MeV. The former result is typical for OSR
\cite{Iud_PN97,Piet_PRC95}. This means that OSR states are
mixtures of $M1(K=1)$ and $E2(K=1)$ modes, which is common in well
deformed nuclei.

Note that, in WFM calculations for $^{164}$Dy
\cite{Bal_EPJC18,Bal_arxiv19,Bal_PAN20},
the lowest $K^{\pi}=1^+$ state at 1.47 MeV has a
huge quadrupole strength $B(E2)$=25.44 W.u. ($\approx$ 1300 e$^2$ fm$^4$).
The authors do not explain origin of this state. Besides, for the next state
at 2.20 MeV, the plots (a-b) of fig. 9 in Ref. \cite{Bal_arxiv19} show a
spurious-like isoscalar flow. By our opinion, the 1.47-MeV state is
spurious,  and higher states can also have spurious admixtures
despite the statements  \cite{Bal_EPJC18,Bal_arxiv19,Bal_PAN20} that spurious
modes are extracted in WFM by construction.
Note that similar lowest-by-energy spurious states appear in QRPA
calculations if the 2qp basis is insufficient and/or the procedure for removal
of spurious states is not exact. Our QRPA calculations for $^{160,162,164}$Dy
(with accurate extraction of spurious admixtures by method \cite{Kva_seEPJA19})
do not give low-energy $K^{\pi}=1^+$ states with so high $B(E2)$, see e.g. Fig. 4.
Moreover, such
states are not known experimentally and, to our knowledge, absent in other
microscopic calculations, see e.g. Ref. \cite{Sol_NPA96} for $^{164}$Dy.

In Table \ref{tab-4}, we show spin, orbital, and total QRPA strengths
$\sum B(M1)$ summed in the SSR (0 - 2.4 MeV), OSR (2.4 - 4 MeV) and
SSR+OSR (0 - 4 MeV) energy intervals.
The total QRPA strengths are compared with NRF experimental data for $1^+$ states
observed at 2.8 - 3.1 MeV in $^{160}$Dy \cite{Wessel_PLB88}),
2.3 - 3.1 MeV in $^{162}$Dy \cite{Mar_exp_NRF_2005} and 2.5 - 3.8 MeV
in $^{164}$Dy  \cite{Mar_exp_NRF_2005}.

Table \ref{tab-4} shows that at 0 - 2.4 MeV the spin strength dominates
over the orbital one. For SkM*, the orbital fraction in this interval is also
essential. In OSR region 2.4 - 4 MeV,  the orbital $M1$ strength strongly dominates
though the spin strength is large as well.

Following Table \ref{tab-4}, QRPA total $M1$ strengths summed at 0-4 MeV
significantly overestimate the experimental values in $^{160,162}$Dy but generally
correspond to the experiment in  $^{164}$Dy (SkM* and SG2). Perhaps, as mentioned
above, the experimental data
for $^{160,162}$Dy \cite{Wessel_PLB88,Mar_exp_NRF_2005} miss a significant part of $M1$
strength. Also, the present calculations do not take into account a coupling with complex
configurations  which can spread the strength and so decrease $\sum B(M1)$-values
at 0-4 MeV.  Our results significantly depend on the applied Skyrme force.
For example, in all considered nuclei,  SVbas gives much smaller orbital and
total strengths than SkM* and SG2. This can be explained by a stronger pairing in SVbas
(see discussion of Table III in Sec. II), which upshifts a part of $M1$ strength above
4 MeV.

In both SSR and OSR regions, we see an interference between spin and orbital
contributions to the total strength (i.e. the sum of spin and orbital contributions
does not equal to  the total strength). It is convenient to estimate this effect
by an interference factor
\begin{equation}
   R=\frac{\sum B(M1)_t}{\sum B(M1)_o + \sum B(M1)_s}
\end{equation}
where $\sum B(M1)_o$,  $\sum B(M1)_s$ and  $\sum B(M1)_t$ are summed orbital,
spin and total strengths. The interference is
destructive at $R<1$, constructive  at $R>1$ and absent at $R=1$.

Table \ref{tab-4} shows that the interference is destructive in SSR range
(with exception of SkM* case in $^{164}$Dy) and constructive in OSR range.
{\it The interference greatly  increases the role of the minor spin
fraction in the OSR range.}
For example, in $^{162}$Dy (SG2), the interference results in the total
strength 4.54 $\mu^2_N$ which is much larger than  the orbital
strength  2.85 $\mu^2_N$.

\begin{table*} % Table V
%\centering
\caption{Characteristics of some relevant low-energy $K^{\pi}_{\nu}=1^+_{\nu}$
states in $^{162}$Dy, calculated  within QRPA with the forces SkM*, SVbas and SG2.
For each state, we show the excitation energy $E$, orbital, spin
and total reduced transition probabilities $B(M1)$ and main 2qp components
(contribution to the state norm  in $\%$, structure in terms of Nilsson
asymptotic quantum numbers, position of the involved single-particle
states relative to the Fermi level F, and original quantum subshells in
the spherical limit).
}
\label{tab-5}       % Give a unique label
% For LaTeX tables you can use
\begin{tabular}{|c|c|c|c|c|c|c|c|c|c|}
\hline
Force & $\nu$ & E &\multicolumn{3}{c|}{$B(M1) \; [\mu^2_N]$} &
\multicolumn{4}{c|}{main 2qp components}
\\
\hline
 &   & [MeV] & \; orb \; & \; spin \; & total & $\%$  & \; $[N,n_z,\Lambda]$ \; &
 F-position  & spher. limit \\
\hline
SkM* & 3 & 1.95 & 0.05 & 0.29 & 0.11 & 69 & pp $[411\uparrow,411\downarrow]$ &
$F-1,F+1$  & $2d_{5/2},2d_{3/2}$ \\
&&&&&& 30 & \; nn $[521\uparrow,521\downarrow]$ \; & \;
$F-1,F+2$ \; & $2f_{7/2},2f_{5/2}$\\
%\hline
& 4 & 2.08 & 0.02 & 0.73 & 0.50 & 69 & nn $[521\uparrow,521\downarrow]$ &
$F-1,F+2$ & $2f_{7/2},2f_{5/2}$\\
&&& &&& 28 & pp $[411\uparrow,411\downarrow]$ &
$F-1,F+1$  & $2d_{5/2},2d_{3/2}$ \\
%\hline
& 8 & 3.09 & 0.86 & 0.05 & 1.33 & 61 & nn $[521\uparrow,512\uparrow]$ &
$F-1,F+4$ & $2f_{7/2},2f_{7/2}$\\
&&& &&& 25 & pp $[411\uparrow,402\uparrow]$ &
$F-1,F+4$  & $2d_{5/2},2d_{5/2}$ \\
\hline
 SVbas  & 1 & 1.88 & 0.05 & 0.54 & 0.27 & 97 &  pp $[411\uparrow,411\downarrow]$ &
 $F,F+1$ & $2d_{5/2},2d_{3/2}$\\
&&&&&& 2 & nn $[521\uparrow,521\downarrow]$ & $F-1,F+2$ & $2f_{7/2},2f_{5/2}$\\
       & 4 & 2.36 & $\sim$0  & 0.20 & 0.18 & 94 & nn $[521\uparrow,521\downarrow]$ &
$F-1,F+2$ & $2f_{7/2},2f_{5/2}$\\
&&&&&& 2 & pp $[411\uparrow,411\downarrow]$ & $F,F+1$ & $2d_{5/2},2d_{3/2}$\\
& 8 & 3.17 & 0.77 & 0.04 & 1.13 & 65 & nn $[521\uparrow,512\uparrow]$ &
$F-1,F+4$ & $2f_{7/2},1h_{9/2}$\\
&&& &&& 16 & pp $[413\downarrow,404\downarrow]$ &
$F-2,F+4$  & $1g_{7/2},1g_{7/2}$ \\
 \hline
 SG2 & 1 & 2.06 & 0.03 & 0.46 & 0.27 & 99 &  pp $[411\uparrow,411\downarrow]$ &
 $F,F+1$ & $2d_{5/2},2d_{3/2}$\\
     & 3 & 2.36 & $\sim$0 &0.26 & 0.28 & 99 & nn $[521\uparrow,521\downarrow]$ &
 $F-1,F+2$ & $2f_{7/2},2f_{5/2}$\\
      & 8 & 3.44 & 0.86 & 0.01 & 1.07 & 57 & nn $[521\uparrow,512\uparrow]$ &
$F-1,F+4$ & $2f_{7/2},1h_{9/2}$\\
&&& &&& 31 & pp $[413\downarrow,404\downarrow]$ &
$F-2,F+4$  & $1g_{7/2},1g_{7/2}$ \\
\hline
\end{tabular}
\\
\end{table*}

Our results generally agree with the study of low-energy (0 - 4 MeV)
$K^{\pi}=1^+$ states in $^{160,162,164}$Dy, performed within
the Quasiparticle-Phonon Nuclear Model (QPNM) \cite{Sol_NPA96}. This model
is not self-consistent. However it has an advantage of taking into account the
coupling with complex configurations. In agreement with our results,
QPNM also predicts in Dy isotopes a well separated group of
$1^+$ states located at 2-2.6 MeV and carrying a noticeable fraction of spin
M1 strength. However, in QPNM the total strength of these states is mainly orbital.
Only in two states at 2.0-2.1 MeV in $^{164}$Dy spin contribution to $M1$ strength
dominates over the orbital one. Coupling with complex configurations is
found strong in OSR region and weaker for lower excitations.
This effect can additionally downshift the orbital strength to the lower SSR region.
QPNM also predicts a considerable interference between spin
and orbital contributions.

For a better understanding of our results, it is worth to consider the structure
and other features of the most interesting $1^+$ states. They are shown for $^{162}$Dy
in Table \ref{tab-5}. We present two states with the largest spin
strength $B(M1)_s$ and one state with the largest orbital strength $B(M1)_o$.
In the spin states, we have $B(M1)_s > B(M1)_o$. Their main 2qp components,
proton $[411\uparrow , 411\downarrow]$ and neutron $[521\uparrow , 521\downarrow]$,
are of the spin-flip character and correspond to particle-hole
(1ph) transitions. Note that the same spin-flip 2qp configurations were found in
QPNM calculations \cite{Sol_NPA96} for low-energy $1^+$ states Dy isotopes.
In the spherical limit, these configurations are reduced to
spin-flip partners $2d_{5/2},2d_{3/2}$  and $2f_{7/2},2f_{5/2}$ with low
orbital moments  $l$=2 and 3. For low $l$, the spin-orbit energy splitting
$\sim (\bf{l}\cdot \bf{s})$ is small and leads to low-energy spin-flip excitations.
The states with larger $l$ contribute to the spin-flip giant resonance
located at a higher energy. Altogether, we see
that {\it so called SSR states are actually ordinary low-energy non-collective
spin-flip excitations}.

\begin{figure} %5
\centering
\includegraphics[width=8cm]{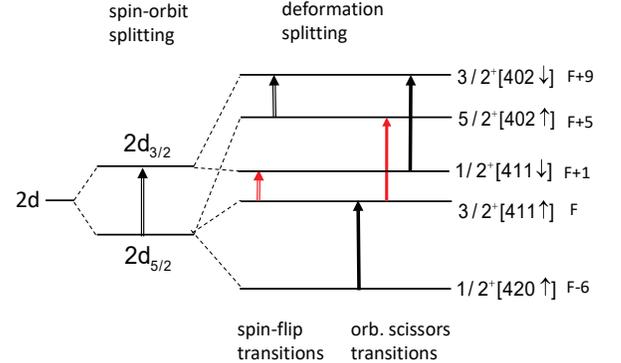}
\caption{A calculated (SG2) scheme of spin-flip (left empty arrows)
and orbital scissors (right filled arrows) $M1$ transitions
 in the proton $2d$ subshell in $^{162}$Dy. As indicated in the top
 inscriptions, the left part of the figure demonstrates a spin-orbit
 splitting into $2d_{3/2}$ and $2d_{5/2}$ levels in the spherical case,
 while the right part exhibits an additional deformation  splitting.
 In the deformed case, $M1(\Delta K)=1$ transitions form two
 groups, spin-flip and orbital scissors, as indicated in the bottom inscriptions.
The Fermi level is $3/2^{+}[411\uparrow]$. The $1ph$ transitions,
spin-flip $3/2^{+}[411\uparrow] \to 1/2^{+}[411\downarrow]$ and orbital
$3/2^{+}[411\uparrow] \to 5/2^{+}[402\uparrow]$, are marked by red color.}
%See more details in the text.}
\label{fig5_2d}
\end{figure}

The orbital and spin-flip $M1$ transitions in $^{162}$Dy can be illustrated using
neutron and proton single-particle level schemes.  In Fig. \ref{fig5_2d}, we show
a proton scheme for $2d$ subshell, calculated with SG2 at the equilibrium deformation
$\beta$=0.346. This scheme demonstrates the same physical mechanisms as in
Fig. \ref{fig2} but now for the case including the proton spin-flip transition
$3/2^+[411\uparrow] \to 1/2^+[411\downarrow]$ which is 
of our interest. We see that the low-energy spin-flip transition
$2d_{5/2} \to 2d_{3/2}$ can take place already in the spherical case.
In the deformed case, two spin-flip and three orbital M1 transitions are possible. However,
only two of these transitions are of $1ph$ character and so not suppressed
(other transitions can appear only  due to the pairing).
They are spin-flip  $3/2^+[411\uparrow] \to 1/2^+[411\downarrow]$  and orbital
$3/2^+[411\uparrow] \to 5/2^+[402\uparrow]$. As seen from
Table \ref{tab-5}, the proton spin-flip 2qp configuration $[411\uparrow , 411\downarrow]$ indeed
dominates in the states at 1.95 MeV (SkM*), 1.88 MeV (SVbas), and 2.06 MeV (SG2).
The orbital configuration $[411\uparrow , 402\uparrow]$ is fragmented
between many states, it is seen e.g. in 3.09-MeV state (SkM*). Since deformations in
$^{160,162,164}$Dy are similar (see Table \ref{tab-2}), the same results
should take place for  $^{160}$Dy and $^{164}$Dy as well.
\begin{figure} %6
\centering
\includegraphics[width=9cm]{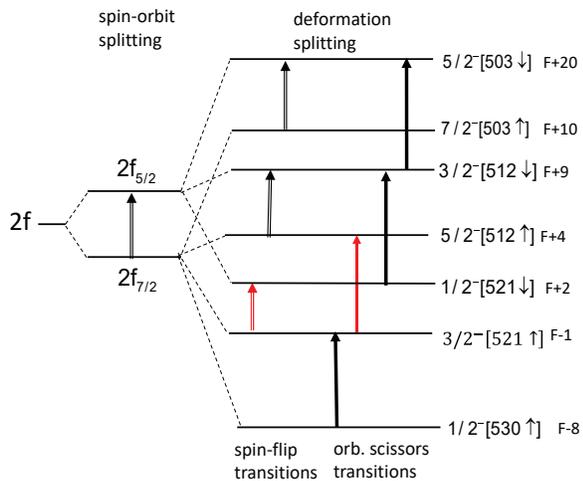}
\caption{The same as in Fig. 5 but for the neutron $2f$ subshell in $^{162}$Dy.
The $1ph$ spin-flip $3/2^{-}[521\uparrow] \to 1/2^{-}[521\downarrow]$
and orbital $3/2^{-}[521\uparrow] \to 5/2^{-}[512\uparrow]$
transitions are marked by red color.}
\label{fig6_2f}
\end{figure}

A similar analysis can be done for a neutron single-particle scheme in $^{162}$Dy.
A relevant part of this scheme for $2f$ subshell is shown in Fig. \ref{fig6_2f}.
We see that again, among many possible
spin-flip and orbital $M1$ transitions, there are only two $1ph$ transitions:
spin-flip $3/2^{-}[521\uparrow] \to 1/2^{-}[521\downarrow]$ and
orbital $3/2^{-}[521\uparrow] \to 5/2^{-}[512\uparrow]$.
The corresponding  2qp configurations  indeed take place in Table \ref{tab-5}.

It is easy to recognize from Fig. \ref{fig6_2f} that $^{160}$Dy and $^{164}$Dy,
whose Fermi levels correspond to $F-1$ and $F+1$ states
of the given  neutron scheme, also allow  1ph spin-flip transitions
$3/2^{-}[521\uparrow] \to 1/2^{-}[521\downarrow]$. This explains
why in our calculations all three isotopes $^{160,162,164}$Dy
demonstrate similar distributions of low-lying spin-flip excitations.

\subsection{Nuclear currents in  $^{162}$Dy}

\begin{figure} %7
\centering
\includegraphics[width=8cm]{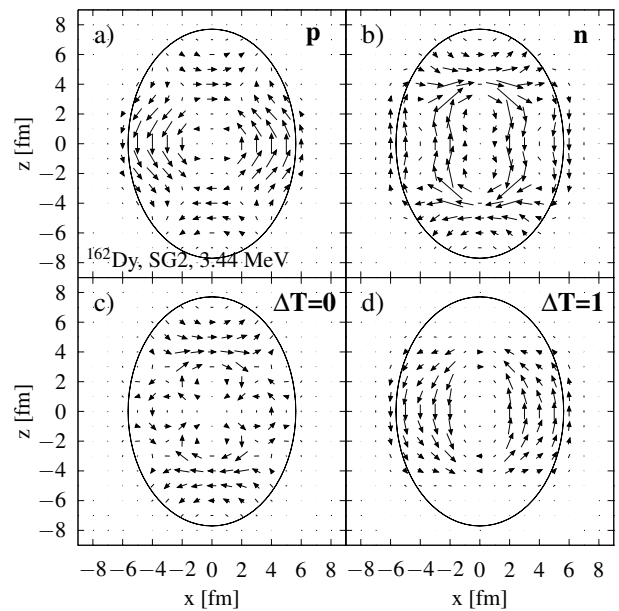}
\caption{Proton (a), neutron (b), isoscalar (c), and isovector (d)
convection CTD in $(x,z)$ plane for 3.44-MeV state in $^{162}$Dy, calculated
within QRPA with the force SG2. A solid ellipse shows the nuclear boundary.}
\label{fig7}
\end{figure}

In this section, we show various CTD $\delta \bold {j}_{\nu} ({\bf r})$
defined in Sec. II. CTD are calculated with the force SG2 for a few relevant
states in $^{162}$Dy, shown in Table \ref{tab-5}. First, we consider
3.44-MeV state which, following Fig.~4,
demonstrates the largest orbital M1 strength. Figure \ref{fig7} shows
for this state the proton, neutron, isoscalar ($\Delta$T=0), and isovector
($\Delta$T=1) CTD plotted on $(x,z)$-plane, where $z$ is the nuclear symmetry
axis. Magnitudes of the currents are equally scaled to provide distinctive
pictures. So, only relative lengths of the current arrows and their
directions (but not absolute lengths of arrows) are matter. The
nuclear boundary estimated for the sharp
nucleus edge is depicted by a solid ellipse.

\begin{figure} %8
\centering
\includegraphics[width=8cm]{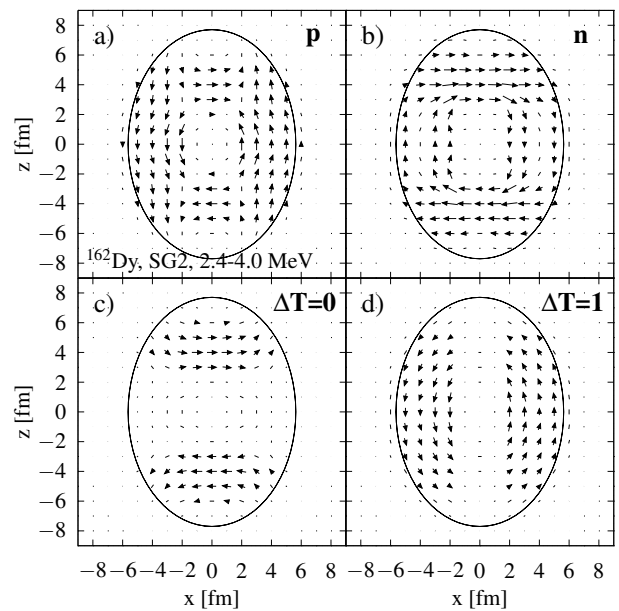}
\caption{The same as in Fig. 7 but for the energy interval 2.4-4 MeV.}
\label{fig8}
\end{figure}

Figure \ref{fig7} shows that protons and neutrons in 3.44-MeV state
move in opposite directions at the left and right surface regions
(cf. plots a), b) and d)) and this motion resembles an isovector OSR
(a similar orbital current was earlier obtained in deformed $^{50}$Cr \cite{Pai16}).
Following Table \ref{tab-5}, 3.44-MeV state has large proton (57$\%$)
and neutron (31$\%$) 2qp components. This complicates a general flow
and makes it different (in the pole regions) from the simple collective
OSR picture. We also see that 3.44-MeV state exhibits both
isoscalar and isovector currents.

For a reasonable comparison with collective WFM currents, it is worth to
consider the {\it summed} CTD involving contributions of all QRPA states
from the OSR energy range 2.4-4 MeV.  The summed CTD  will smooth
individual peculiarities of the currents  of particular QRPA states and thus
highlight the main (e.g. collective) features of the nuclear flow in the given
energy range. The procedure to get summed CTD is described in Ref.
\cite{Rep_PRC13}. The summed CTD are shown in Fig. \ref{fig8}. The
flow in left/right surface regions  now more resembles the OSR picture.
However, the flow is again mixed by isospin. It is isovector in the
left/right sides and isoscalar in the pole regions.

\begin{figure} %9
\centering
\includegraphics[width=8.2cm]{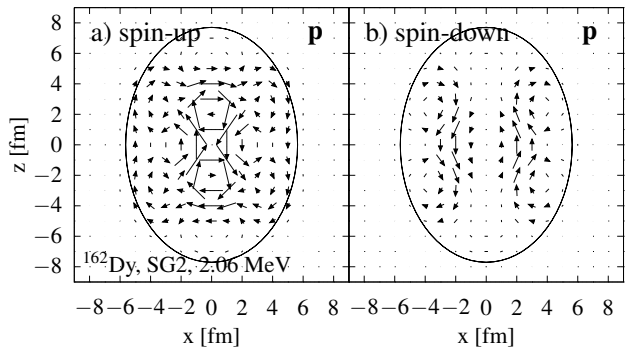}
\caption{Proton spin-up (a) and spin-down (b) CTD
in  basically proton 2.06-MeV state in $^{162}$Dy.}
\label{fig9}
\end{figure}
\begin{figure} %10
\centering
\includegraphics[width=8.2cm]{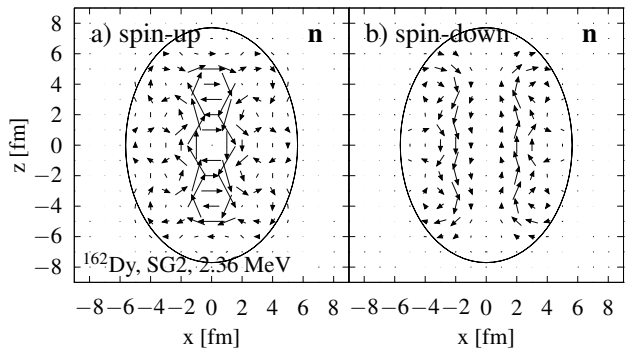}
\caption{The same as in Fig. 9 but for CTD in basically neutron
2.36-MeV state in $^{162}$Dy.}
\label{fig10}
\end{figure}

Following WFM \cite{Bal_arxiv19,Bal_PAN20}, the low-energy spin states should
demonstrate out-of-phase rotation-like oscillations of spin-up and
spin-down nuclear fractions, see Fig. 1 b). To check this prediction,
we show in Figs. \ref{fig9} and \ref{fig10} spin-up and spin-down CTD
for spin-flip states at 2.06 and 2.36 MeV. As seen from
Table \ref{tab-5}, these states are almost fully exhausted by one proton
and one neutron 2qp component, respectively. So, to characterize the nuclear flow in
these states, the corresponding proton and neutron spin-up and
spin-down currents are enough.  Figures \ref{fig9} and \ref{fig10}
show that the currents are not regular but rather demonstrate a complex cellular-like
structure formed by the dominant 2qp configurations. They are proton
$[411\uparrow, 411\downarrow]$ and neutron $[521\uparrow, 521\downarrow]$ configurations
arising from $2d (l=2)$ and $2f (l=3)$ spherical subshells. Accordingly,
the proton  flow in Fig. \ref{fig9} has a fewer number of cells than the
neutron one in Fig. \ref{fig10}.

\begin{figure} %11
\centering
\includegraphics[width=8.2cm]{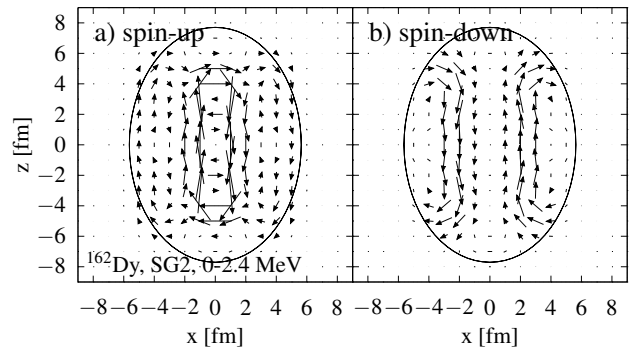}
\caption{Spin-up (a) and spin-down (b) CTD for the energy interval 0-2.4 MeV}
\label{fig11}
\end{figure}

In Fig. \ref{fig11}, the summed CTD are depicted. They do not match regular
collective  WFM spin-scissors currents shown in Ref. \cite{Bal_arxiv19}.

\begin{figure} %12
\centering
\includegraphics[width=8.5cm]{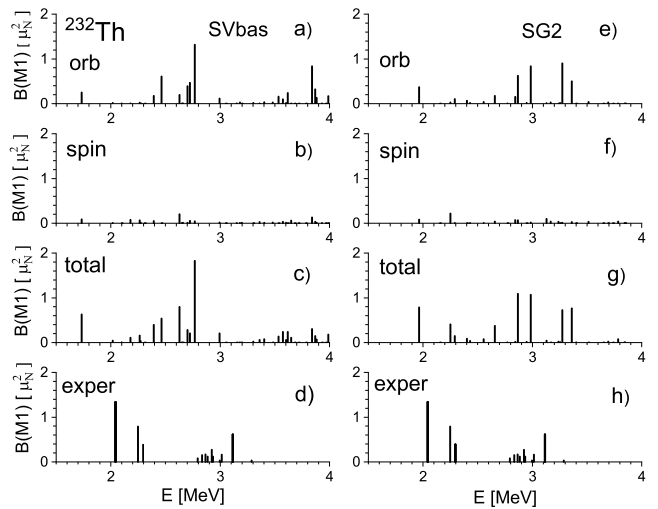}
\caption{The computed (SVbas, SG2) orbital, spin  and total
low-energy M1 strengths in $^{232}$Th
 as compared with the experimental data \cite{Ade_232Th}.}
\label{fig12}
\end{figure}

\begin{table*} % Table VII
%\centering
\caption{The same as in Table \ref{tab-5} but for states in $^{232}$Th.}
%For each state, the maximal 2qp component is shown.}
\label{tab-6}       % Give a unique label
% For LaTeX tables you can use
\begin{tabular}{|c|c|c|c|c|c|c|c|c|c|}
\hline
Force &$\nu$ & E &\multicolumn{3}{c|}{$B(M1) \; [\mu^2_N]$} &
\multicolumn{4}{c|}{main 2qp components}
\\
\hline
& & [MeV] & orb & spin & total & $\%$  & $[N,n_z,\Lambda]$ &
 F-position  & spher. limit \\
\hline
& 1 & 1.73 &\; 0.25 \;& \; 0.09 \; & 0.63 \;& 91 & \; pp $[660\uparrow, 651\uparrow]$ \; &
$F,F+1$  & $1i_{13/2},1i_{13/2}$ \\
%&&&&&& 2 & pp $[532\downarrow,523\downarrow]$ &
%$F-3,F+2$ & $1h_{9/2},1h_{9/2}$\\
%\hline
SVbas & 15 & 2.77 & 1.32 & 0.04 & 1.82 & 51 & nn $[761\uparrow, 752\uparrow]$ &
%$[633\downarrow, 624\downarrow]$ &
 $F-1,F-3$ & $1j_{15/2},1j_{15/2}$\\
%& &&&&& 30 & pp $[532\downarrow,523\downarrow]$ &
%$F-3,F+2$ & $1h_{9/2},1h_{9/2}$\\
\hline
& 1 & 1.96 &\; 0.18 \;& \; 0.04 \; & 0.39 \;& 89 & \; pp $[660\uparrow, 651\uparrow]$ \; &
$F,F+1$  & $1i_{13/2},1i_{13/2}$ \\
%&&&&&& 3 & nn $[624\downarrow,633\downarrow]$ &
%$F-3,F+2$ & $2g_{9/2},2g_{9/2}$\\
%\hline
SG2 & 3 & 2.25 & 0.01 & 0.11 & 0.20 & 96 & nn $[631\uparrow, 631\downarrow]$ &
 $F,F+3$ & $1i_{13/2},3d_{5/2}$\\
%& &&&&& 30 & pp $[532\downarrow,523\downarrow]$ &
%$F-3,F+2$ & $1h_{9/2},1h_{9/2}$\\
 & 13 & 2.98 & 0.42 & 0.01 & 0.53 & 32 & pp  $[530\uparrow,521\uparrow]$ &
$F-1,F+4$ & $2f_{7/2},2f_{7/2}$\\
%& &&&&& 21 & nn $[624\downarrow,633\downarrow]$ &
%$F-3,F+2$ & $2g_{9/2},2g_{9/2}$\\
\hline
\end{tabular}
\\
\end{table*}

\subsection{$M1$ strength in $^{232}$Th}

In addition to strongly deformed Dy isotopes, SSR was also predicted by WFM
in a less deformed nucleus $^{232}$Th \cite{Bal_PRC18,Mol_EPJC18,Bal_arxiv19}.
In this nucleus, the experiment \cite{Ade_232Th} also gives two separate
groups of low-energy $1^+$ states (see plot (d) in Fig. \ref{fig12}).
The lower group at $E <$ 2.5 MeV is considered by WFM as a candidate
for SSR. In this connection, we present QRPA results for $^{232}$Th, obtained
with the forces SVbas and SG2.
Note that these forces, especially SG2, provide a good description of
the spin-flip M1 giant resonance in $^{232}$Th, see Appendix A.

\begin{table} % Table VII
\centering
\caption{The computed orbital, spin and total $B(M1)$ strengths
summed at E=0-3.3 MeV as compared with the experimental data \cite{Ade_232Th}.
$R$ are the interference factors.}
\label{tab-7}       % Give a unique label
\begin{tabular}{|c|c|c|c|c|c|}
\hline
 Force & \multicolumn{4}{|c|}{$\sum B(M1)  [\mu_N^2]$} & R\\
\hline
 & orb & spin & total  & exper & \\
\hline
 SVbas & 3.60 & 0.66 & 5.23 &     & 1.23
\\
SG2 & 3.37 & 0.68 & 4.92 & 4.26 & 1.21
\\
\hline
\end{tabular}
\end{table}
In Fig. \ref{fig12}, the computed orbital, spin, and total $B(M1)$
strengths in $^{232}$Th are compared with NRF experimental data \cite{Ade_232Th}.
We see that the spin strength is much smaller that the orbital
one even at $E < $  2.5 MeV.  For SG2, there is a remarkable agreement
between the distribution of the total strength and the experimental data.
Namely, both experiment and theory give at $E <$ 2.5 MeV the distinctive
group of the states. Fig. \ref{fig12} obviously {\it does not demonstrate
any distinctive SSR}. Indeed, both level groups, below and above 2.5 MeV,
are strongly dominated by the orbital strength. So, {\it these two groups
are explained not by separation of SSR and OSR modes (as was suggested
by WFM)  but rather by a fine structure of the OSR alone.}

In Table \ref{tab-6}, we show the features of some representative
states with the large spin and orbital strength. In SVbas case, the first state is not
spin-flip despite it has the largest spin strength at the range $E <$  2.5 MeV.
Moreover, it is dominated by the orbital strength.
This state is not collective and demonstrates a constructive interference
of the spin and orbital contributions, in contrast to the lowest states in Dy isotopes.
The 2.77-MeV state  is collective and exhibits a constructive interference like the
orbital states in Dy case. In SG2 case, the first 1.96-MeV state is non-collective
and mainly orbital (like for SVbas).  The third 2.98-MeV state is spin-flip one with
the dominant neutron  configuration $[631\uparrow, 631\downarrow]$. Both states
demonstrate a strong constructive interference of orbital and spin contributions.
The 2.98-MeV state is a collective orbital state.

The calculated and experimental summed $M1$ strengths  are compared in Table \ref{tab-7}.
Like in Dy isotopes,  the theoretical values of the total $\sum B(M1)$  somewhat
overestimate the experimental data. As mentioned in Sec. III-A, the overestimation
can be caused  by i) missing of a significant part of $M1$ strength in the experiment
and ii) neglect of the coupling with complex configurations. Like in Dy isotopes,
we see in $^{232}$Th the constructive interference of the spin and orbital contributions
to the total strength.

\section{Conclusions}

The WFM prediction of a low-energy spin-scissors resonance (SSR) in deformed nuclei
\cite{Bal_NPA11,Bal_PRC18,Mol_EPJC18,Bal_EPJC18,Bal_arxiv19,Bal_PAN20}
was analyzed in the framework of the self-consistent QRPA
approach using Skyrme forces SkM*, SVbas, and SG2.
The calculations were performed for deformed nuclei $^{160,162,164}$Dy and
$^{232}$Th. Two of these nuclei, $^{164}$Dy and
$^{232}$Th, were proposed by WFM as promising candidates for SSR.

The calculations have shown that, in strongly deformed nuclei like
$^{160,162,164}$Dy, indeed there can exist a group of $K^{\pi}=1^+$
spin states located at 1.5-2.4 MeV, i.e. below the conventional
orbital scissor resonance (OSR). Following our analysis, these states
are ordinary spin-flip excitations characterized by $M1(\Delta K =1)$
transitions between spin-orbit partners in subshells with a low orbital
momentum $l$, e.g. $2d$ and $1f$. Such low-$l$ spin-flip
states can form a separate low-energy group if a large deformation
shifts OSR to a higher energy. In our calculations, this is the
case for well deformed $^{160,162,164}$Dy
but  not for less deformed $^{232}$Th.

The obtained  low-energy spin states are non-collective and mainly
exhausted by one 2qp spin-flip configuration. This can be explained by
basically isovector character of the spin-spin
residual interaction which upshifts the collectivity to higher energies.
The non-collective character of low-energy spin states contradicts with the
collective scissors nature of the predicted SSR. Further, the calculated distributions
of nuclear currents locally resemble the OSR collective flow but not the SSR one.

Since OSR energy $E\approx 66 \delta A^{-1/3}$ MeV falls with the
mass number A,
this resonance in heavy (actinide) nuclei goes down by energy and mixes
with nearby spin states.  Being stronger, OSR conceals these states.
So heavy deformed nuclei are not suitable to exhibit distinctive
low-energy spin states.

At the excitation energy $E < $ 4 MeV, most of $1^+$ states demonstrate a
significant interference of spin-flip and orbital contributions to
$M1$ strength. The interference considerably increases the
total $M1$  strength in the OSR energy range. This should be taken into
account while comparing the computed strengths with estimations derived
merely for the orbital mode. A part of the orbital strength is downshifted to
the region of spin states ($E \le $ 2.4 MeV) and, vice versa, the OSR
region hosts some spin-flip strength.

The experimental data \cite{Mar_exp_NRF_2005,Val_exp_2015,Ren_exp_Oslo_2018,Ade_232Th}
show two distinctive low-energy groups of $1^+$ states in $^{162,164}$Dy and
$^{232}$Th. These two groups are treated by WFM as SSR and OSR. Our calculations show
that lowest $1^+$ states in $^{160,162,164}$Dy are indeed of spin-flip character.
However they are located  at $E \le $ 2.4 MeV, i.e. below the observed states.
So perhaps both two  observed groups are produced by
fragmentation of the orbital strength. This is even more the case in $^{232}$Th where
the low-energy spin strength is almost negligible. So, by our opinion,
{\it the available experimental data still do not confirm  the existence of SSR}.
 More definite conclusions can be drawn after further experimental and theoretical effort.
Indeed, following discussion \cite{Ren_exp_Oslo_2018}, a significant number of $1^+$ states
can be found below 2.7 MeV, see database \cite{database}  for candidates.
As for the theory, it should take into account the coupling  with complex configurations,
which, in principle, can redistribute the $M1$ strength.

The WFM {\it scissor-like} treatment of low-energy spin M1 excitations
requires the nuclear deformation. In other words, spin-scissors excitations
can exist only in deformed nuclei. Instead, our calculations
show that low-energy spin states arise from the spin-orbit splitting and so
can exist even in spherical nuclei. So {\it the deformation is not the
origin of the low-energy spin strength} but only an essential factor affecting
its properties. In principle, WFM  does not use any two-rotor assumption.
Then, perhaps, the deformation-induced scissors-like scheme
is just a poorly chosen illustration.

The spin-orbit splitting and spin-spin residual interaction are of a primary
importance in the exploration of spin excitations \cite{Ves_PRC09,Nes_JPG10}.
To check the accuracy of our QRPA method in
description of these factors, we performed calculations  for the spin-flip M1(K =1)
giant resonance in $^{162}$Dy and $^{232}$Th  and obtained for the forces SVbas and SG2
a good agreement with the experiment. The same test should be done by WFM as well.

In WFM calculations  \cite{Bal_EPJC18,Bal_arxiv19,Bal_PAN20}, the lowest
$K^{\pi}=1^+$ state
with the energy E=1.47 MeV  has a huge quadrupole strength B(E2)=25.4 W.u..
The authors do not explain the origin of such state. By our opinion, this state
is spurious. Neither experimental  data, nor our QRPA calculations
for $^{160,162,164}$Dy and $^{232}$Th give at E$<$ 4 MeV  $1^+$ states
with so large B(E2) value.

The discrepancy between WFM and QRPA predictions for spin states in $^{232}$Th
could be clarified by $(p,p')$ measurements which are sensitive to spin-flip
excitations and not so much to orbital ones. If low-energy spin states indeed
exist in $^{232}$Th, they should be observed in $(p,p')$ reaction.

Since low-energy spin states are reduced to almost pure 2qp excitations, these states
can be useful for investigation of low-$l$ spin-orbit splitting and its interplay
with nuclear deformation. Besides, such states can be also useful for testing tensor forces.

\section*{Acknowledgement}

We thank Profs. P.-G. Reinhard, P. von Neumann-Cosel and A.V. Sushkov
for useful discussions.  The work was partly supported
by Votruba - Blokhintsev (Czech Republic - BLTP JINR) grant (VON and JK) and
a grant of the Czech Science Agency, Project No. 19-14048S (JK). VON and
WK appreciate the Heisenberg-Landau grant (Germany - BLTP JINR).
A.R. acknowlegdes the support by the Slovak Research and Development
Agency under contract No. APVV-15-0225, Slovak grant agency VEGA
(contract No. 2/0067/21), and the Research and Development Operational
Programme funded by the European Regional
Development Fund, project  No. ITMS code 26210120023.

\appendix

\section{$M1$ spin-flip giant resonance}
% in $^{162}$Dy and $^{232}$Th}
\label{app:A}

Energy and structure of $M1$ spin-flip giant resonance  in open-shell nuclei
are basically determined  by the interplay between  spin-orbital splitting in
proton and neutron schemes from one side and spin-spin residual interaction from
another side \cite{Hei10,Ves_PRC09,Nes_JPG10}.
To check the accuracy of our approach, we present here QRPA results for spin-flip
giant resonance in $^{162}$Dy and $^{232}$Th,
obtained with the Skyrme parameterizations  SkM*, SVbas, and SG2.
We were not able to find experimental data for this
resonance in $^{162}$Dy. So, for this nucleus, we compare QRPA results with the
$(p,p')$ data for the neighbouring nucleus $^{158}$Gd \cite{Fre90}
which has a similar quadrupole deformations ($\beta_2$=0.348) \cite{database}.
For $^{232}$Th, we use $(p,p')$ data \cite{238U,Sarr96}.

\begin{figure} %13
\centering
\includegraphics[width=8.5cm]{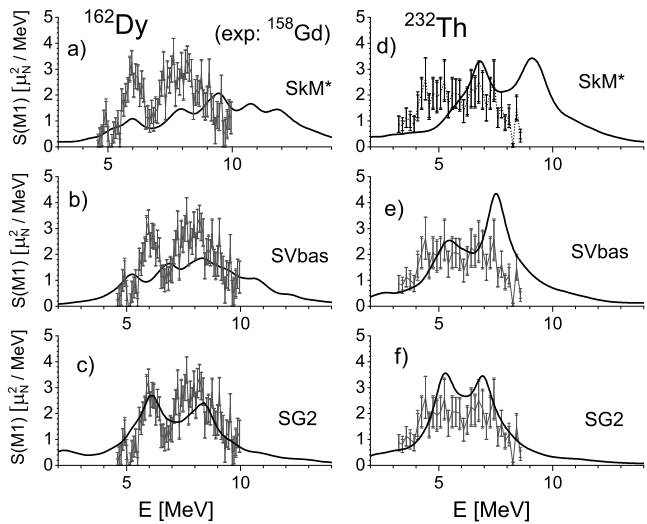}
\caption{$M1$ spin-flip giant resonance in $^{162}$Dy and $^{232}$Th,
calculated with Skyrme forces SkM*, SVbas and SG2. The results
are compared with scaled experimental data (in arbitrary units) for
$^{232}$Th \cite{238U,Sarr96} (right plots) and
neighbouring nucleus  $^{158}$Gd \cite{Fre90} (left plots). See details in the text.}
\label{fig13}
\end{figure}

In Fig. \ref{fig13}, the results of our calculations are compared with the experimental
data. QRPA strength functions are obtained by averaging transition rates
$B_{\nu}(M1)$ for separate QRPA states  by Lorentz weight with an averaging parameter
$\Delta$=1 MeV, see Refs. \cite{Ves_PRC09,Nes_JPG10} for more detail. Only spin part of M1
transition operator (\ref{eq:M1}) is used. The experimental data (in arbitrary units)
are properly scaled for a convenient comparison with QRPA strength functions.
Fig. \ref{fig13} shows that SVbas, and especially SG2, well describe localization and fine structure
of the resonance in both nuclei. In SkM*, distribution of the strength is too wide and
upshifted to higher energies. This difference can be explained by smaller values of spin-flip
parameters $b_4$ and $b'_4$ in SVbas and SG2 sets (see Table I in Sec. II).

\begin{table} % Table VIII
\centering
\caption{The strength $B(M1)_s$ summed at E=0-12 MeV in our SkM*,
SVbas and SG2 calculations  as compared with QRPA (SG2)
results of Sarriguren et al \cite{Sarr96}.}
\label{tab-8}       % Give a unique label
\begin{tabular}{|c|c|c|c|c|}
\hline
Nucleus & \multicolumn{4}{c|}{$\sum B(M1)_s [\mu_N^2]$}\\
\hline
        &  SkM* & SVbas & SG2 & Sarriguren \cite{Sarr96}\\
\hline
$^{160}$Dy  & \; 14.5 & 13.2 & 12.4 & 11.4  \\
\hline
$^{162}$Dy  & \; 14.7 & 13.4 & 12.7  &  12.2  \\
\hline
$^{164}$Dy  & \; 14.7 & 13.6 & 12.9  & 12.2 \\
\hline
$^{232}$Th &  \; 17.2 & \; 15.9 & \: 14.3 & 14.9 \\
\hline
\end{tabular}
\end{table}

In Table \ref{tab-8}, the spin $B(M1)$-values
%reduced transition probabilities
summed at the energy interval E =0-12 MeV are compared with early
QRPA results of P. Sarriguren et al \cite{Sarr96}, obtained
with the force SG2. It is seen that the agreement is fine for SG2,
acceptable for SVbas and  worse for SkM*.

Altogether, Fig. \ref{fig13} and Table \ref{tab-8} show that forces SVbas
and SG2 are most relevant for exploration of spin-flip excitations.

\section{WFM vs QRPA results}

In this Appendix, we briefly discuss some important points
concerning the comparison and treatment of WFM and QRPA results.

It is known that macroscopic  and microscopic models often successfully
supplement each other in description of nuclear modes \cite{Har01,RSbook}.
For example, our QRPA results for isovector E1 giant resonance \cite{Kle_PRC08},
E1 toroidal mode \cite{Rep_PRC13,Rep_EPJA19} and  M1 orbital scissors (present
calculations) well agree
with predictions of macroscopic models  \cite{GT48,SJ50}, \cite{Sem81},
and \cite{Iud78,Iud_PN97,Iud_NC00}, respectively.  However, we were
not able to get a similar correspondence between our QRPA results
and WFM predictions for SSR.
In this connection, it is worth to discuss some important issues.

1) {\it Accuracy of WFM  numerical results for $M1$ low-energy spin states.}

Both Skyrme QRPA and WFM have spin-orbit mean-field terms and so include
spin-orbit splitting and corresponding spin-flip excitations. In QRPA, spin-flip
states are identified by strong domination of spin-flip 2qp components, large values of
spin $B(M1)_s$, and hindered $B(E2)$. WFM deals with collective variables and identifies
spin states mainly by enhanced $B(M1)$ and hindered $B(E2)$, where $B(M1)$ is calculated
only for the total (spin+orbital) M1 operator. This seems not enough to identify
reliably spin-flip states. Besides, following Table I for $^{164}$Dy in
\cite{Bal_arxiv19}, WFM does not produce at all the M1 spin-flip giant resonance,
which makes questionable the accuracy of WFM in description of spin-flip states.
Further, the parameters of the WFM Hamiltonian (includes
a spherical harmonic  oscillator, spin-orbit terms, pairing,
quadrupole-quadrupole  and spin-spin separable residual interaction)
are taken from different sources and, by our opinion, not properly justified.
In this connection, the claimed good agreement of WFM results with the experimental
data looks doubtful.

2) {\it Is the spin-scissors scheme generally relevant?}

The SSR macroscopic picture was suggested in analogy with OSR scheme
developed within the two-rotor model \cite{Iud78,Iud_PN97,Iud_NC00}.
However, the OSR scheme was confirmed by experimentally  observed
\cite{Zie_PRL90,Iud_PLB93} specific dependencies of OSR energy and
strength on the nuclear deformation \cite{Hei10,Iud_PN97,Iud_NC00}.
Instead, the WFM calculations have not still
suggested any specific measurable features justifying the relevance of the
spin-scissors  picture.

The spin-scissors picture assumes a non-zero nuclear deformation. Without
deformation this picture cannot be realized in principle. However,
following our calculations, the deformation is not the primary origin of
$M1$ low-energy spin states.

In the spin-scissors picture (Fig. 1(b,c)), SSR looks
as a two-step process including spin-flip excitation + orbital
oscillation. It is not clear how to match such two-step process with
the linear regime used in WFM. We have not found in Refs.
\cite{Bal_PRC18,Mol_EPJC18,Bal_EPJC18,Bal_arxiv19,Bal_PAN20} any
relevant linear probe external field to generate such SSR.

Following Eq. (29) in Ref. \cite{Bal_arxiv19},
the WFM nuclear current is formed solely by components of an orbital collective
variables for different combinations of spin directions. Maybe, for this reason,
the currents for OSR, SSR-I and SSR-II  in Figs. 9-11 of Ref. \cite{Bal_arxiv19} look
identical (up to direction of the motion). Our QRPA distributions of the nuclear
current partly support the isovector OSR scheme but not the SSR one.

Altogether, we have a feeling that the deformation-induced scissors-like
picture used for illustration and interpretation of the WFM results is
a poor and even misleading choice.

\section{Matrix elements of magnetic transitions
in axially deformed nuclei}

In cylindrical coordinates, the single-particle wave function
with quantum numbers $K^{\pi}$  has the spinor form
\begin{equation}
\label{eq:psi_i_spin}
\Psi_{i}(\bold r) =
\left(\begin{array}{c} {R_{i}^{(+)}(\rho,z) e^{im_{i}^{(+)}\phi}} \\
{R_{i}^{(-)}(\rho,z) e^{im_{i}^{(-)}\phi}} \end{array}
 \right) ,
\end{equation}
for the normal state and
\begin{equation}
\label{eq:psi_i_bar_spin}
\Psi_{\overline{i}}(\bold r) = \hat T \Psi_{i}(\bold r) =
\left(\begin{array}{c}
{-R_{i}^{(-)}(\rho,z) e^{-im_{i}^{(-)}\phi}} \\
{R_{i}^{(+)}(\rho,z) e^{-im_{i}^{(+)}\phi}} \end{array}
 \right)
\end{equation}
for the time-reversal state. Here the momentum projection is decomposed as
$K_i=m_{i}^{(\sigma)}+\frac{1}{2} \sigma$
%\end{equation}
with $ \sigma = \pm 1$.

The spin and orbital  $M\lambda\mu$ transition operators are \cite{BMv1}
\begin{eqnarray}\label{eq:MLam}
\hat{S}_{l \lambda \mu} &=& \mu_N \sqrt{\lambda (2\lambda +1)}r^{l}
 g^{q}_s \{ {\hat{\bold s}} \; Y_{l}\}_{\lambda\mu} \; ,
%+(-1)^{\mu}\{\sigma Y_{\lambda-1}\}_{\lambda -\mu})
\\
 \hat{L}_{l \lambda \mu} &=& \mu_N \sqrt{\lambda (2\lambda +1)}r^{l}
 g^{q}_l \frac{2}{\lambda +1}\{{\hat{\bold l}}\; Y_{l}\}_{\lambda\mu}
\end{eqnarray}
where $l=\lambda-1$, $\mu_N$ is the nuclear magneton, $\hat{\bold s}$
and $\hat{\bold l}$ are standard spin and orbital operators,
$g^{q}_s$ and $g^{q}_l$ are spin and orbital gyromagnetic factors.
Further
\begin{eqnarray}
\{ \hat{\bold s}\; Y_{l} \}_{\lambda\mu} &=&
  \sum_{m}\sum_{\alpha=-1,0,1} C^{\lambda \mu}_{l m, 1 \alpha} Y_{lm} {\hat s}_{\alpha},
\\
\{ \hat{\bold l}\; Y_{l} \}_{\lambda\mu} &=&
  \sum_{m}\sum_{\alpha=-1,0,1} C^{\lambda \mu}_{l m, 1 \alpha} Y_{lm} {\hat l}_{\alpha}
\end{eqnarray}
where $Y_{lm}$ are the spherical harmonics and  $C^{\lambda \mu}_{l m, 1 \alpha}$
are Clebsch-Gordan coefficients.

The matrix elements for the orbital and spin $M\lambda\mu$ transitions
from the BCS vacuum $|\textrm{BCS}\rangle$ to the two-quasiparticle (2qp) state
$\alpha^+_i\alpha^+_{\bar j}|\textrm{BCS}\rangle$ with the selection rule
$|K_i-K_j|=\mu$ ($K_i, K_j >0,  \mu \ge 0$) have the form
\begin{eqnarray}
\label{tiLbarj}
&& \langle i\bar{j} | \hat{L}_{l \lambda \mu} | \textrm{BCS} \rangle
=  2 \pi \mu_N  \sqrt{\lambda (2\lambda +1)} \frac{2g_l}{\lambda +1}
  u^{(-)}_{ij}
 \\
  \nonumber
 &\cdot& \int dz d\rho \rho
  \{
  g_{l\mu}\; C^{\lambda\mu}_{l\mu, 10} \;
 [ R^{(+)}_i m^{(+)}_{j} R^{(+)}_{j}
 + R^{(-)}_i m^{(-)}_{j} R^{(-)}_{j}]
 \\
 \nonumber
 &+& \frac{1}{\sqrt{2}} g_{l \mu+1}\; C^{\lambda \mu}_{l\mu+1, 1-1}
 [ R^{(+)}_i (\rho\frac{d}{dz} - z \frac{d}{d\rho}
 - m^{(+)}_{j} \frac{z}{\rho}) R^{(+)}_{j}
 \\
 \nonumber
 && \qquad\qquad\qquad\quad
 + R^{(-)}_i  (\rho\frac{d}{dz} - z \frac{d}{d\rho}
 - m^{(-)}_{j} \frac{z}{\rho})R^{(-)}_{j}]
 \\
 \nonumber
  &+& \frac{1}{\sqrt{2}} g_{l \mu-1}\; C^{\lambda \mu}_{l\mu-1, 1 1}
 [ R^{(+)}_i (\rho\frac{d}{dz} - z \frac{d}{d\rho}
 + m^{(+)}_{j} \frac{z}{\rho}) R^{(+)}_{j}
\\
 \nonumber
 && \qquad\qquad\qquad\quad
 + R^{(-)}_i  (\rho\frac{d}{dz} - z \frac{d}{d\rho}
  + m^{(-)}_{j} \frac{z}{\rho})R^{(-)}_{j}]
  \}\; ,
\end{eqnarray}
%and
\begin{eqnarray}
\label{tiSbarj}
&& \langle i \bar{j}| \hat{S}_{l \lambda \mu} | \textrm{BCS}\rangle = 2 \pi \mu_N
   \sqrt{\lambda (2\lambda +1)} g_s
   u^{(-)}_{ij}
\\
 \nonumber
 &\cdot&\int dz d\rho  \rho
 \; \{
 \frac{1}{2} g_{l\mu}\; C^{\lambda\mu}_{l\mu, 10} \;
 [ R^{(+)}_i R^{(+)}_j - R^{(-)}_i  R^{(-)}_j]
 \\
 \nonumber
 &&\qquad\quad + \frac{1}{\sqrt{2}} g_{l \mu+1}\;
   C^{\lambda \mu}_{l\mu+1, 1-1}  R^{(-)}_i R^{(+)}_j
 \\
 \nonumber
 &&\qquad\quad - \frac{1}{\sqrt{2}} g_{l \mu-1}\;
    C^{\lambda \mu}_{l\mu-1, 1 1} R^{(+)}_i R^{(-)}_j
 \} .
\end{eqnarray}
Here $ u^{(-)}_{ij}=u_iv_j-u_jv_i$ is the combination of Bogoliubov factors.
The $(\rho, z)$- dependence in the functions
$R^{(\pm)}_i$,  $g_{l\mu}$ and $g_{l\mu\pm 1}$ is omitted for the sake of
simplicity.

For the selection rule $K_i+K_j=\mu$, the matrix elements for the transitions
to the 2qp state $\alpha^+_i\alpha^+_j|\textrm{BCS}\rangle$ read
\begin{eqnarray}
\label{tiLj}
&& \langle ij | \hat{L}_{l \lambda \mu} | \textrm{BCS} \rangle
=  2 \pi \mu_N  \sqrt{\lambda (2\lambda +1)} \frac{2g_l}{\lambda +1}
  u^{(-)}_{ij}
 \\
  \nonumber
 &\cdot&\int dz d\rho \rho \;
  \{
  g_{l\mu}\; C^{\lambda\mu}_{l\mu, 10} \;
 [R^{(-)}_i m^{(+)}_{j} R^{(+)}_{j}- R^{(+)}_i m^{(-)}_{j} R^{(-)}_{j}]
 \\
 \nonumber
 &+& \frac{1}{\sqrt{2}} g_{l \mu+1}\; C^{\lambda \mu}_{l\mu+1, 1-1}
 [ R^{(+)}_i (\rho\frac{d}{dz} - z \frac{d}{d\rho}
 + m^{(-)}_j \frac{z}{\rho}) R^{(-)}_j
 \\
 \nonumber
 && \qquad\qquad\qquad\quad
 - R^{(-)}_i  (\rho\frac{d}{dz} - z \frac{d}{d\rho}
 + m^{(+)}_j \frac{z}{\rho})R^{(+)}_j]
 \\
 \nonumber
  &+& \frac{1}{\sqrt{2}} g_{l \mu-1}\; C^{\lambda \mu}_{l\mu-1, 1 1}
 [ R^{(+)}_i (\rho\frac{d}{dz} - z \frac{d}{d\rho}
 - m^{(-)}_{j} \frac{z}{\rho}) R^{(-)}_{j}
\\
 \nonumber
 && \qquad\qquad\qquad\quad
 - R^{(-)}_i  (\rho\frac{d}{dz} - z \frac{d}{d\rho}
  - m^{(+)}_j \frac{z}{\rho})R^{(+)}_j]
  \} \; ,
\end{eqnarray}
\begin{eqnarray}
\label{tiSj}
&& \langle ij| \hat{S}_{l \lambda \mu} | \textrm{BCS}\rangle
= 2 \pi \mu_N \sqrt{\lambda (2\lambda +1)} g_s u^{(-)}_{ij}
\\
 \nonumber
 &\cdot&\int dz d\rho \rho
 \; \{
 \frac{1}{2} g_{l\mu}\; C^{\lambda\mu}_{l\mu, 10} \;
 [ R^{(+)}_i R^{(-)}_j + R^{(-)}_i  R^{(+)}_j]
 \\
 \nonumber
 &&\qquad\quad + \frac{1}{\sqrt{2}} g_{l \mu+1}\;
   C^{\lambda \mu}_{l\mu+1, 1-1}  R^{(-)}_i R^{(-)}_j
 \\
 \nonumber
 &&\qquad\quad + \frac{1}{\sqrt{2}} g_{l \mu-1}\;
    C^{\lambda \mu}_{l\mu-1, 1 1} R^{(+)}_i R^{(+)}_j
 \} .
\end{eqnarray}
In (\ref{tiLbarj})-(\ref{tiSj}), the functions $g_{lm}$
($m=\mu, \mu\pm 1)$ are
\begin{equation}
g_{lm}(\rho, z)= r^l Y_{lm}(\theta, \phi) e^{-im\phi} .
\end{equation}

In the case of our interest ($\lambda\mu=11$), the transition operator has the form
(\ref{eq:M1}). In the above expressions, we have $l=0$   and so only the terms with
$\mu-1=0$ survive. In these terms, $g_{l\mu-1}(\rho, z) \to g_{00} = 1/\sqrt{4\pi}$ and
finally we get
\begin{eqnarray}
\label{tiLbarj11}
&& \langle i\bar{j} | \hat{L}_{011} | \textrm{BCS} \rangle
=  \sqrt{\frac{3\pi}{2}}\mu_N g_l u^{(-)}_{ij}
%  \int dz \int \rho d\rho
 \\
  \nonumber
 &\cdot&\int dz d\rho \rho \;
 [ R^{(+)}_i (\rho\frac{d}{dz} - z \frac{d}{d\rho}
 + m^{(+)}_{j} \frac{z}{\rho}) R^{(+)}_{j}
\\
 \nonumber
 && \quad\qquad\quad
 + R^{(-)}_i  (\rho\frac{d}{dz} - z \frac{d}{d\rho}
  + m^{(-)}_{j} \frac{z}{\rho})R^{(-)}_{j}] \; ,
\end{eqnarray}
\begin{equation}
\label{tiSbarj11}
\langle i \bar{j}| \hat{S}_{011} | \textrm{BCS}\rangle
= - \sqrt{\frac{3\pi}{2}}\mu_N g_s u^{(-)}_{ij}
\int dz d\rho  \rho R^{(+)}_i R^{(-)}_j \; ,
\end{equation}
\begin{eqnarray}
\label{tiLj11}
&& \langle ij | \hat{L}_{011} | \textrm{BCS} \rangle
= \sqrt{\frac{3\pi}{2}}\mu_N g_l u^{(-)}_{ij}
 \\
  \nonumber
 &\cdot&\int dz d\rho \rho \;
 [ R^{(+)}_i (\rho\frac{d}{dz} - z \frac{d}{d\rho}
 - m^{(-)}_{j} \frac{z}{\rho}) R^{(-)}_{j}
\\
 \nonumber
 && \quad\qquad\quad
 - R^{(-)}_i  (\rho\frac{d}{dz} - z \frac{d}{d\rho}
  - m^{(+)}_j \frac{z}{\rho})R^{(+)}_j]
  \} \; ,
\end{eqnarray}
\begin{equation}
\label{tiSj11}
\langle ij| \hat{S}_{011} |\textrm{BCS} \rangle
= \sqrt{\frac{3\pi}{2}}\mu_N g_s u^{(-)}_{ij}
\int dz d\rho \rho
  R^{(+)}_i R^{(+)}_j \; .
\end{equation}

\end{document}